# Active Clustering: Robust and Efficient Hierarchical Clustering using Adaptively Selected Similarities

Brian Eriksson, Gautam Dasarathy, Aarti Singh, Robert Nowak*

**Abstract**

Hierarchical clustering based on pairwise similarities is a common tool used in a broad range of scientific applications. However, in many problems it may be expensive to obtain or compute similarities between the items to be clustered. This paper investigates the hierarchical clustering of $N$ items based on a small subset of pairwise similarities, significantly less than the complete set of $N(N-1)/2$ similarities. First, we show that if the intracluster similarities exceed intercluster similarities, then it is possible to correctly determine the hierarchical clustering from as few as $3N \log N$ similarities. We demonstrate this order of magnitude savings in the number of pairwise similarities necessitates sequentially selecting which similarities to obtain in an adaptive fashion, rather than picking them at random. We then propose an *active clustering* method that is robust to a limited fraction of anomalous similarities, and show how even in the presence of these noisy similarity values we can resolve the hierarchical clustering using only $O\left(N \log^2 N\right)$ pairwise similarities.

## 1 Introduction

Hierarchical clustering based on pairwise similarities arises routinely in a wide variety of engineering and scientific problems. These problems include inferring gene behavior from microarray data [1], Internet topology discovery [2], detecting community structure in social networks [3], advertising [4], and database management [5, 6]. It is often the case that there is a significant cost associated with obtaining each similarity value. For example, in the case of Internet topology inference, the determination of similarity values requires many probe packets to be sent through the network, which can place a significant burden on the network resources. In other situations, the similarities may be the result of

*B. Eriksson is with the Department of Computer Science, Boston University. G. Dasarathy and R. Nowak are with the Department of Electrical and Computer Engineering, University of Wisconsin - Madison. A. Singh is with the Machine Learning Department, Carnegie Mellon University.

expensive experiments or require an expert human to perform the comparisons, again placing a significant cost on their collection.

The potential cost of obtaining similarities motivates a natural question: Is it possible to reliably cluster items using less than the complete, exhaustive set of all pairwise similarities? We will show that the answer is yes, particularly under the condition that intracluster similarity values are greater than intercluster similarity values, which we will define as the *Tight Clustering* (TC) condition. We also consider extensions of the proposed approach to more challenging situations in which a significant fraction of intracluster similarity values may be smaller than intercluster similarity values. This allows for robust, provably-correct clustering even when the TC condition does not hold uniformly.

The TC condition is satisfied in many situations. For example, the TC condition holds if the similarities are generated by a branching process (or tree structure) in which the similarity between items is a monotonic increasing function of the distance from the branching root to their nearest common branch point (ancestor). This sort of process arises naturally in clustering nodes in the Internet [7]. Also notice that, for suitably chosen similarity metrics, the data can satisfy the TC condition even when the clusters have complex structures. For example, if similarities between two points are defined as the length of the longest edge on the shortest path between them on a nearest-neighbor graph, then they satisfy the TC condition given the clusters do not overlap. Additionally, density based similarity metrics [8] also allow for arbitrary cluster shapes while satisfying the TC condition.

One natural approach is to try to cluster using a small subset of randomly chosen pairwise similarities. However, we show that this is quite ineffective in general. We instead propose an *active* approach that sequentially selects similarities in an adaptive fashion, and thus we call the procedure *active clustering*. We show that under the TC condition, it is possible to reliably determine the unambiguous hierarchical clustering of $N$ items using at most $3N \log N$ of the total of $N(N-1)/2$ possible pairwise similarities. Since it is clear that we must obtain at least one similarity for each of the $N$ items, this is about as good as one could hope to do. Then, to broaden the applicability of the proposed theory and method, we propose a robust active clustering methodology for situations where a random subset of the pairwise similarities are unreliable and therefore fail to meet the TC condition. In this case, we show how using only $O\left(N \log^2 N\right)$ actively chosen pairwise similarities, we can still recover the underlying hierarchical clustering with high probability.

Both of the clustering methodologies rely solely on the relative ordering between the similarities, which means they are invariant to strictly monotonic transformations of the similarities (*e.g.,* scaling or shifts). Therefore, these techniques will be automatically robust without the need for an additional preprocessing step for situations where similarity calibration is an issue, such as subjective human-annotated features arising in applications like the *Netflix Problem* [9].

While there have been prior attempts at developing robust procedures for hierarchical clustering [10, 11, 12], these works do not try to optimize the number of similarity values needed to robustly identify the true clustering, and mostly require all $O\left(N^2\right)$ similarities. Other prior work has attempted to develop efficient active clustering methods [13, 14, 15], but the proposed techniques are ad-hoc and do not provide any theoretical guarantees. Outside of clustering literature there are some interesting connections emerge between this problem and prior work on graphical model inference [2, 16], which we exploit here.

The paper is organized as follows. The hierarchical clustering problem and our tight clustering condition is introduced in Section 2. In Section 3, we describe the proposed methodology for resolving the hierarchical clustering using a limited number of pairwise values in the noiseless setting. Robust methods for clustering in the presence of similarity errors and outliers are derived in Section 4. Finally, experiments on synthetic and real data sets are presented in Section 5.

## 2 The Hierarchical Clustering Problem

Let $\mathbf{X} = \{x_1, x_2, \ldots, x_N\}$ be a collection of $N$ items. Our goal will be to resolve a *hierarchical clustering* of these items.

**Definition 1.** *A **cluster** $\mathcal{C}$ is defined as any subset of $\mathbf{X}$. A collection of clusters $\mathcal{T}$ is called a **hierarchical clustering** if $\cup_{\mathcal{C}_i \in \mathcal{T}} \mathcal{C}_i = \mathbf{X}$ and for any $\mathcal{C}_i, \mathcal{C}_j \in \mathcal{T}$, only one of the following is true **(i)** $\mathcal{C}_i \subset \mathcal{C}_j$, **(ii)** $\mathcal{C}_j \subset \mathcal{C}_i$, **(iii)** $\mathcal{C}_i \cap \mathcal{C}_j = \emptyset$.*

The hierarchical clustering $\mathcal{T}$ has the form of a tree, where each node corresponds to a particular cluster. The tree is *binary* if for every $\mathcal{C}_k \in \mathcal{T}$ that is not a leaf of the tree, there exists proper subsets $\mathcal{C}_i$ and $\mathcal{C}_j$ of $\mathcal{C}_k$, such that $\mathcal{C}_i \cap \mathcal{C}_j = \emptyset$, and $\mathcal{C}_i \cup \mathcal{C}_j = \mathcal{C}_k$. The binary tree is said to be *complete* if it has $N$ leaf nodes, each corresponding to one of the individual items. Without loss of generality, we will assume that $\mathcal{T}$ is a complete (possibly unbalanced) binary tree, since any non-binary tree can be represented by an equivalent binary tree.

Let $\mathbf{S} = \{s_{i,j}\}$ denote the collection of all pairwise similarities between the items in $\mathbf{X}$, with $s_{i,j}$ denoting the similarity between $x_i$ and $x_j$ and assuming $s_{i,j} = s_{j,i}$. The traditional hierarchical clustering problem uses the complete set of pairwise similarities to infer $\mathcal{T}$. In order to guarantee that $\mathcal{T}$ can be correctly identified from $\mathbf{S}$, the similarities must conform to the hierarchy of $\mathcal{T}$. We consider the following sufficient condition.

**Definition 2.** *The triple $(\mathbf{X}, \mathcal{T}, \mathbf{S})$ satisfies the **Tight Clustering (TC) Condition** if for every set of three items $\{x_i, x_j, x_k\}$ such that $x_i, x_j \in \mathcal{C}$ and $x_k \notin \mathcal{C}$, for some $\mathcal{C} \in \mathcal{T}$, the pairwise similarities satisfies, $s_{i,j} > \max(s_{i,k}, s_{j,k})$.*

In words, the TC condition implies that the similarity between all pairs within a cluster is greater than the similarity with respect to any item outside the cluster. We can consider using off-the-shelf hierarchical clustering methodologies, such as bottom-up agglomerative clustering [17], on the set of pairwise similarities that satisfies the TC condition. Bottom-up agglomerative clustering is a recursive process that begins with singleton clusters (*i.e.,* the $N$ individual items to be clustered). At each step of the algorithm, the pair of most similar clusters are merged. The process is repeated until all items are merged into a single cluster. It is easy to see that if the TC condition is satisfied, then the standard bottom-up agglomerative clustering algorithms such as single linkage, average linkage and complete linkage will all produce $\mathcal{T}$ given the complete similarity matrix **S**. Various agglomerative clustering algorithms differ in how the similarity between two clusters is defined, but every technique requires all $N(N-1)/2$ pairwise similarity values since all similarities must be compared at the very first step.

To properly cluster the items using fewer similarities requires a more sophisticated adaptive approach where similarities are carefully selected in a sequential manner. Before contemplating such approaches, we first demonstrate that adaptivity is necessary, and that simply picking similarities at random will not suffice.

**Proposition 1.** *Let $\mathcal{T}$ be a hierarchical clustering of $N$ items and consider a cluster of size $m$ in $\mathcal{T}$ for some $m \ll N$. If $n$ pairwise similarities, with $n < \frac{N}{m}(N-1)$, are selected uniformly at random from the pairwise similarity matrix $S$, then any clustering procedure will fail to recover the cluster with high probability.*

*Proof.* In order for any procedure to identify the $m$-sized cluster, we need to measure at least $m-1$ of the $\binom{m}{2}$ similarities between the cluster items. Let $p = \binom{m}{2}/\binom{N}{2}$ be the probability that a randomly chosen similarity value will be between items inside the cluster. If we uniformly sample $n$ similarities, then the expected number of similarities between items inside the cluster is approximately $n\binom{m}{2}/\binom{N}{2}$ (for $m \ll N$). Given Hoeffding's inequality, with high probability the number of observed pairwise similarities inside the cluster will be close to the expected value. It follows that we require $n\binom{m}{2}/\binom{N}{2} = n\frac{m(m-1)}{N(N-1)} \geq m-1$, and therefore we require $n \geq \frac{N}{m}(N-1)$ to reconstruct the cluster with high probability. □

This result shows that if we want to reliably recover clusters of size $m = N^\alpha$ (where $\alpha \in [0,1]$), then the number of randomly selected similarities must exceed $N^{(1-\alpha)}(N-1)$. In simple terms, randomly chosen similarities will not adequately sample all clusters. As the cluster size decreases (*i.e.,* as $\alpha \to 0$) this means that almost all pairwise similarities are needed if chosen at random. This is more than are needed if the similarities are selected in a sequential and adaptive manner. In Section 3, we propose a sequential method that requires at most $3N \log N$ pairwise similarities to determine the correct hierarchical clustering.

# 3 Active Hierarchical Clustering under the TC Condition

From Proposition 1, it is clear that unless we acquire almost all of the pairwise similarities, reconstruction of the clustering hierarchy when sampling at random will fail with high probability. In this section, we demonstrate that under the assumption that the TC condition holds, an *active clustering* method based on *adaptively selected* similarities enables one to perform hierarchical clustering efficiently. Towards this end, we consider the work in [16] where the authors are concerned with a very different problem, namely, the identification of causality relationships among binary random variables. We present a modified adaptation of prior work here in the context of our hierarchical clustering from pairwise similarities problem.

From our discussion in the previous section, it is easy to see that the problem of reconstructing the hierarchical clustering $\mathcal{T}$ of a given set of items $X = \{x_1, x_2, \ldots, x_N\}$ can be reinterpreted as the problem of recovering a binary tree whose leaves are $\{x_1, x_2, \ldots, x_N\}$. In [16], the authors define a special type of test on triples of leaves called the *leadership test* which identifies the "leader" of the triple in terms of the underlying clustering tree structure. A leaf $x_k$ is said to be the *leader* of the triple $(x_i, x_j, x_k)$ if the path from the root of the tree to $x_k$ does not contain the nearest common ancestor of $x_i$ and $x_j$. An example of this property can be seen in Figure 1. This prior work shows that one can efficiently reconstruct the entire tree $\mathcal{T}$ using only these leadership tests.

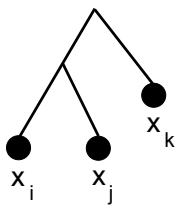

Figure 1: Tree structure where $x_k$ is the leader of the triple $(x_i, x_j, x_k)$.

The following lemma demonstrates that given observed pairwise similarities satisfying the TC condition, an `outlier` test using pairwise similarities will correctly resolve the leader of a triple of items.

**Lemma 1.** *Let $\mathbf{X}$ be a collection of items equipped with pairwise similarities $\mathbf{S}$ and hierarchical clustering $\mathcal{T}$. For any three items $\{x_i, x_j, x_k\}$ from $\mathbf{X}$, define*

$$\texttt{outlier}(x_i, x_j, x_k) = \begin{cases} x_i : \max(s_{i,j}, s_{i,k}) < s_{j,k} \\ x_j : \max(s_{i,j}, s_{j,k}) < s_{i,k} \\ x_k : \max(s_{i,k}, s_{j,k}) < s_{i,j} \end{cases} \qquad (1)$$

*If $(\mathbf{X}, \mathcal{T}, \mathbf{S})$ satisfies the TC condition, then $\texttt{outlier}(x_i, x_j, x_k)$ coincides with the leader of the same triple with respect to the tree structure conveyed by $\mathcal{T}$.*

*Proof.* Suppose that $x_k$ is the leader of the triple with respect to $\mathcal{T}$. This occurs if and only if there is a cluster $\mathcal{C} \in \mathcal{T}$ such that $x_i, x_j \in \mathcal{C}$ and $x_k \in \mathcal{T} \setminus \mathcal{C}$. By the TC condition, this implies that $s_{i,j} > \max(s_{i,k}, s_{j,k})$. Therefore $x_k$ is the `outlier` of the same triple. □

Note that `outlier` relies only on the ordering of similarity values, and therefore it is invariant to monotonic transformations of the similarities. More precisely, let $f$ be a strictly monotonic function and define $f(\mathbf{S}) := \{f(s_{i,j})\}$ to be the set of pairwise similarities under this transformation. If $(\mathbf{X}, \mathcal{T}, \mathbf{S})$ satisfies the TC condition, then so does $(\mathbf{X}, \mathcal{T}, f(\mathbf{S}))$. In words, the TC condition does not require precise calibration of similarity values.

The clustering algorithm we propose is called `OUTLIERcluster`, and is given below in Algorithm 1. The procedure is based on the `outlier` test and a tree reconstruction algorithm due to [16]. In Theorem 3.1, we show that the algorithm determines the correct hierarchical clustering using $O(N \log N)$ pairwise similarities.

**Theorem 3.1.** *Assume that the triple $(\mathbf{X}, \mathcal{T}, \mathbf{S})$ satisfies the Tight Clustering (TC) condition where $\mathcal{T}$ is a complete (possible unbalanced) binary tree that is unknown. Then, `OUTLIERcluster` recovers $\mathcal{T}$ exactly using at most $3N \log_{3/2} N$ adaptively selected pairwise similarity values.*

*Proof.* From Appendix II of [16], we find a methodology that requires at most $N \log_{3/2} N$ leadership tests to exactly reconstruct the unique hierarchical clustering of $N$ items. Lemma 1 shows that under the TC condition, each leadership test can be performed using only 3 adaptively selected pairwise similarities. Therefore, we can reconstruct the hierarchical clustering $\mathcal{T}$ from a set of items $\mathbf{X}$ using at most $3N \log_{3/2} N$ adaptively selected pairwise similarity values. □

### 3.1 Tight Clustering Experiments

In Table 1 we see the results of both clustering techniques (`OUTLIERcluster` and bottom-up agglomerative clustering) on various synthetic tree topologies given the Tight Clustering (TC) condition. The performance is in terms of the number of pairwise similarities required by the agglomerative clustering methodology, denoted by $n_{agg}$, and the number of similarities required by our `OUTLIERcluster` method, $n_{outlier}$. The methodologies are performed on both a balanced binary tree of varying size ($N = 128, 256, 512$) and a synthetic Internet tree topology generated using the technique from [18]. As seen in the table, our technique resolves the underlying tree structure using at most 11% of the pairwise similarities required by the bottom-up agglomerative clustering approach. As the number of items in the topology increases, further improvements are seen using `OUTLIERcluster`. Due to the pairwise similarities satisfying the TC condition, both methodologies resolve a binary representation of the underlying tree structure exactly.

**Algorithm 1** - OUTLIERcluster($\mathbf{X}, \mathbf{S}$)

**Given :**

1. set of items, $\mathbf{X} = \{x_1, x_2, ...x_N\}$.
2. matrix of pairwise similarities, $\mathbf{S}$.

**Clustering Process :**

Initialize clustering tree $\mathcal{T} = \{x_1, x_2, \{x_1, x_2\}\}$.

**For** $i = \{3, 4, ..., N\}$

1. Set tree $\mathcal{T}_i = \mathcal{T}$.
2. **While** number of items in $\mathcal{T}_i$, denoted $\#\mathcal{T}_i$, is greater than 2
   (a) Select a subtree, $\mathcal{T}_C \subset \mathcal{T}_i$, such that the number of items in the subtree, $\#\mathcal{T}_C$, satisfies $\frac{\#\mathcal{T}_i}{3} < \#\mathcal{T}_C \leq \frac{2\#\mathcal{T}_i}{3}$.
   (b) Find items $x_j, x_k \in \mathcal{T}_C$ such that no subtrees in $\mathcal{T}_C$ (besides $\mathcal{T}_C$) contain both $x_j, x_k$.
   (c) **If:** $x_i = \text{outlier}(x_i, x_j, x_k)$, replace $\mathcal{T}_C$ in $\mathcal{T}_i$ with item $x_j$, **Else:** set $\mathcal{T}_i = \mathcal{T}_C$.
3. Let $x_j, x_k$ be the two remaining items in $\mathcal{T}_i$.
4. Let $\mathcal{T}'$ be the smallest subtree in $\mathcal{T}$ containing both $x_j, x_k$. This subtree contains $\mathcal{T}'_j, \mathcal{T}'_k$, such that $x_j \in \mathcal{T}'_j$, $x_k \in \mathcal{T}'_k$, $\mathcal{T}' = \mathcal{T}'_j \bigcup \mathcal{T}'_k$, and $\mathcal{T}'_j \bigcap \mathcal{T}'_k = \emptyset$.
5. **If:** $\text{outlier}(x_i, x_j, x_k) = x_i$, replace $\mathcal{T}'$ in $\mathcal{T}$ with $\{\mathcal{T}', \{x_i\}\}$.
   **Elseif:** $\text{outlier}(x_i, x_j, x_k) = x_j$, replace $\mathcal{T}'_k$ in $\mathcal{T}$ with $\{\mathcal{T}'_k, \{x_i\}\}$.
   **Else:** $\text{outlier}(x_i, x_j, x_k) = x_k$, replace $\mathcal{T}'_j$ in $\mathcal{T}$ with $\{\mathcal{T}'_j, \{x_i\}\}$.

**Output :** Hierarchical cluster tree, $\mathcal{T}$.

Table 1: Comparison of OUTLIERcluster and Agglomerative Clustering on various topologies satisfying the Tight Clustering condition.

| Topology | Size | $n_{agg}$ | $n_{outlier}$ | $\frac{n_{outlier}}{n_{agg}}$ |
|---|---|---|---|---|
| Balanced Binary | $N = 128$ | 8,128 | 876 | 10.78% |
| | $N = 256$ | 32,640 | 2,206 | 6.21% |
| | $N = 512$ | 130,816 | 4,561 | 3.49% |
| Synthetic Internet | $N = 768$ | 294,528 | 8,490 | 2.88% |

## 3.2 Fragility of OUTLIERcluster

OUTLIERcluster determines the correct clustering hierarchy when all the pairwise similarities are consistent with the hierarchy $\mathcal{T}$, but it can fail if one or more of the pairwise similarities are inconsistent. In Figure 2, we examine the performance of OUTLIERcluster on a balanced binary tree (with $N = 256$) when a small number of calls to outlier in OUTLIERcluster return an incorrect item with respect to the underlying hierarchy $\mathcal{T}$. The incorrect cases are selected at random. This performance measured in terms of $r_{min}$, the size of the smallest correctly resolved cluster (*i.e.,* all clusters of size $r_{min}$ or larger are reconstructed correctly for the clustering) averaged over 150 separate experiments. As seen in the table, with only two outlier tests erroneous at random, we find that this corrupts the clustering reconstruction using OUTLIERcluster significantly. This can attributed to the greedy construction of the clustering hierarchy using this methodology, where if one of the initial items is incorrectly placed in the hierarchy, this will result in a cascading effect that will drastically reduce the accuracy the clustering.

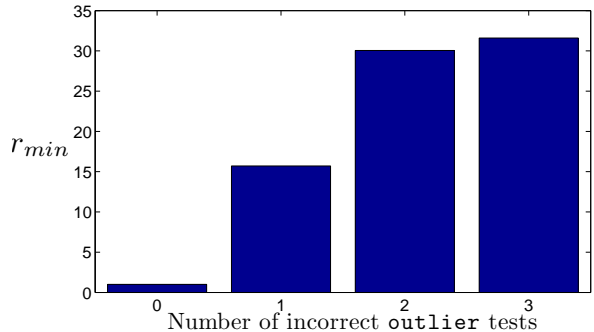

Figure 2: Fragility of OUTLIERcluster when a selected number of outlier tests are incorrect for a balanced binary tree of size $N = 256$.

## 4 Robust Active Clustering

Suppose that most, *but not all*, of the outlier tests agree with $\mathcal{T}$. This may occur if a subset of the similarities are in some sense inconsistent, erroneous or anomalous. We will assume that a certain subset of the similarities produce correct outlier tests and the rest may not. These similarities that produce correct tests are said to be *consistent* with the hierarchy $\mathcal{T}$. *Our goal is to recover the clusters of $\mathcal{T}$ despite the fact that the similarities are not always consistent with it.*

**Definition 3.** *The subset of consistent similarities is denoted $\mathbf{S}_C \subset \mathbf{S}$. These similarities satisfy the following property: if $s_{i,j}, s_{j,k}, s_{i,k} \in \mathbf{S}_C$ then* outlier$(x_i, x_j, x_k)$ *returns the leader of the triple $(x_i, x_j, x_k)$ in $\mathcal{T}$ (i.e., the outlier test is consistent with respect to $\mathcal{T}$).*

We adopt the following probabilistic model for $\mathbf{S}_C$. Each similarity in $\mathbf{S}$ fails to be consistent independently with probability at most $q < 1/2$ (*i.e.,* membership in $\mathbf{S}_C$ is termed by repeatedly tossing a biased coin). The expected cardinality of $\mathbf{S}_C$ is $\mathbb{E}[|\mathbf{S}_C|] \geq (1-q)|\mathbf{S}|$. Under this model, there is a large probability that one or more of outlier tests will yield an incorrect leader with respect to $\mathcal{T}$. Thus, our tree reconstruction algorithm in Section 3 will fail to recover the tree with large probability. We therefore pursue a different approach based on a top-down recursive clustering procedure that uses voting to overcome the effects of incorrect tests.

The key element of the top-down procedure is a robust algorithm for correctly splitting a given cluster in $\mathcal{T}$ into its two subclusters, presented in Algorithm 1. Roughly speaking, the procedure quantifies how frequently two items tend to agree on outlier tests drawn from a small random subsample of other items. If they tend to agree frequently, then they are clustered together; otherwise they are not. We show that this algorithm can determine the correct split of the input cluster $\mathcal{C}$ with high probability. The degree to which the split is "balanced" affects performance, and we need the following definition.

**Definition 4.** *Let $\mathcal{C}$ be any non-leaf cluster in $\mathcal{T}$ and denote its subclusters by $\mathcal{C}_L$ and $\mathcal{C}_R$; i.e., $\mathcal{C}_L \bigcap \mathcal{C}_R = \emptyset$ and $\mathcal{C}_L \bigcup \mathcal{C}_R = \mathcal{C}$. The* balance factor *of $\mathcal{C}$ is $\eta_\mathcal{C} := \min\{|\mathcal{C}_L|, |\mathcal{C}_R|\} \backslash |\mathcal{C}|$.*

**Theorem 4.1.** *Let $0 < \delta' < 1$ and threshold $\gamma \in (0, 1/2)$. Consider a cluster $\mathcal{C} \in \mathcal{T}$ with balance factor $\eta_\mathcal{C} \geq \eta$ and disjoint subclusters $\mathcal{C}_R$ and $\mathcal{C}_L$, and assume the following conditions hold:*

- ***A1** - The pairwise similarities are consistent with probability at least $1-q$, for some $q \leq 1 - \frac{1}{\sqrt{2(1-\delta')}}$.*

- ***A2** - $q$, $\eta$ satisfy $(1 - (1-q)^2) < \gamma < (1-q)^2 \eta$.*

*If $m \geq c_0 \log(4n/\delta')$ and $n > 2m$ (where the constant $c_0$ depends on $q$, $\gamma$, $\eta$ and $\delta'$), then with probability at least $1 - \delta'$ the output of* split$(\mathcal{C}, \mathtt{m}, \delta')$ *is the correct subclusters, $\mathcal{C}_R$ and $\mathcal{C}_L$.*

The proof of the theorem is given in the Appendix. The theorem above shows that the algorithm is guaranteed (with high probability) to correctly split clusters that are sufficiently large for a certain range of $q$ and $\eta$, as specified by **A2**. A bound on the constant $c_0$ is given in Equation 3 in the proof, but the important fact is that it does not depend on $n$, the number of items in $\mathcal{C}$. Thus all but the very smallest clusters can be reliably split. Note that total number of similarities required by split is at most $3mn$. So if we take $m = c_0 \log(4n/\delta')$, the total is at most $3c_0 n \log(4n/\delta')$. The key point of the lemma is this: *instead of using all $O(n^2)$ similarities,* split *only requires $O(n \log n)$*.

The allowable range in **A2** is non-degenerate and covers an interesting regime of problems in which $q$ is not too large and $\eta$ is not too small, this is shown in Figure 3. The allowable range of $\gamma$ cannot be determined without knowledge of $\eta$ and $q$, so in practice $\gamma \in (0, 1/2)$ is a user-selected parameter (we use $\gamma = 0.30$ in all our experiments in the following section), and the Theorem holds for the corresponding set of $(q, \eta)$ in **A2**.

**Algorithm 2** : `split(`$\mathcal{C}, \mathtt{m}, \gamma$`)`

**Input :**

1. A single cluster $\mathcal{C}$ consisting of $n$ items.

2. Parameters $m < n/2$ and $\gamma \in (0, 1/2)$

**Initialize :**

1. Select two subsets $\mathcal{S}_V, \mathcal{S}_A \subset \mathcal{C}$ uniformly at random (with replacement) containing $m$ items each.

2. Select a "seed" item $x_j \in \mathcal{C}$ uniformly at random and let $\mathcal{C}_j \in \{\mathcal{C}_R, \mathcal{C}_L\}$ denote the subcluster it belongs to.

**Split :**

- For each $x_i \in \mathcal{C}$ and $x_k \in \mathcal{S}_A \setminus x_i$, compute the *outlier fraction* of $\mathcal{S}_V$:

$$c_{i,k} := \frac{1}{|\mathcal{S}_V \setminus \{x_i, x_k\}|} \sum_{x_\ell \in \mathcal{S}_V \setminus \{x_i, x_k\}} \mathbf{1}_{\{\mathtt{outlier}(x_i, x_k, x_\ell) = x_\ell\}}$$

where $\mathbf{1}$ denotes the indicator function.

- Compute the *outlier agreement* on $\mathcal{S}_A$:

$$a_{i,j} := \sum_{x_k \in \mathcal{S}_A \setminus \{x_i, x_j\}} \left( \mathbf{1}_{\{c_{i,k} > \gamma \text{ and } c_{j,k} > \gamma\}} + \left( \mathbf{1}_{\{c_{i,k} < \gamma \text{ and } c_{j,k} < \gamma\}} \right) / |\mathcal{S}_A \setminus \{x_i, x_j\}| \right.$$

- Assign item $x_i$ to a subcluster according to

$$x_i \in \begin{cases} \mathcal{C}_j & : \text{if} \quad a_{i,j} \geq 1/2 \\ \mathcal{C}/\mathcal{C}_j & : \text{if} \quad a_{i,j} < 1/2 \end{cases}$$

**Output :** subclusters $\mathcal{C}_j, \mathcal{C}/\mathcal{C}_j$.

We now give our robust active hierarchical clustering algorithm, `RAcluster`. Given an initial single cluster of $N$ items, the `split` methodology of Algorithm 1 is recursively performed until all subclusters are of size less than or equal to $2m$, the minimum resolvable cluster size where we can overcome inconsistent similarities through voting. The output is a hierarchical clustering $\mathcal{T}'$. The algorithm is summarized in Algorithm 3.

Theorem 4.1 shows that it suffices to use $O(n \log n)$ similarities for each call of `split`, where $n$ is the size of the cluster in each call. Now if the splits are balanced, the depth of the complete cluster tree will be $O(\log N)$, with $O(2^\ell)$ calls to `split` at level $\ell$ involving clusters of size $n = O(N/2^\ell)$. An easy calcu-

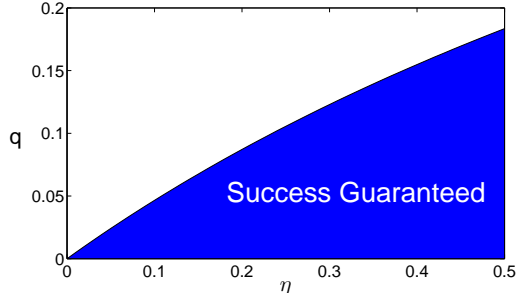

Figure 3: The shaded region depicts the range of $q$ and $\eta$ for which Theorem 4.1 can guarantee correct recovery of subclusters using the `split` algorithm.

---

**Algorithm 3** : `RAcluster`$(\mathcal{C}, \mathtt{m}, \gamma)$

**Given :**

1. $\mathcal{C}$, $n$ items to be hierarchically clustered.
2. parameters $m < n/2$ and $\gamma \in (0, 1/2)$

**Partitioning :**

1. Find $\{\mathcal{C}_L, \mathcal{C}_R\} = \mathtt{split}(\mathcal{C}, \mathtt{m}, \gamma)$.

2. Evaluate hierarchical subtrees, $\mathcal{T}_L, \mathcal{T}_R$, of cluster $\mathcal{C}$ using:

$$\mathcal{T}_L = \begin{cases} \mathtt{RAcluster}(\mathcal{C}_L, \mathtt{m}, \gamma) & : \quad \text{if } |\mathcal{C}_L| > 2m \\ \mathcal{C}_L & : \quad \text{otherwise} \end{cases}$$

$$\mathcal{T}_R = \begin{cases} \mathtt{RAcluster}(\mathcal{C}_R, \mathtt{m}, \gamma) & : \quad \text{if } |\mathcal{C}_R| > 2m \\ \mathcal{C}_R & : \quad \text{otherwise} \end{cases}$$

**Output :** Hierarchical clustering $\mathcal{T}' = \{\mathcal{T}_L, \mathcal{T}_R\}$ containing subclusters of size $\geq 2m$.

---

lation then shows that the total number of similarities required by `RAcluster` is then $O(N \log^2 N)$, compared to the total number which is $O(N^2)$. The performance guarantee for the robust active clustering algorithm are summarized in the following main theorem.

**Theorem 4.2.** *Let $\mathbf{X}$ be a collection of $N$ items with underlying hierarchical clustering structure $\mathcal{T}$ and let $0 < \delta < 1$. If $m = k_0 \log\left(\frac{8}{\delta} N\right)$, for a constant $k_0 > 0$, then `RAcluster`$(\mathbf{X}, m, \gamma)$ uses $O(N \log^2 N)$ similarities and with probability at least $1 - \delta$ recovers all clusters $\mathcal{C} \in \mathcal{T}$ that have size $> 2m$, with balance factor $\eta_\mathcal{C} \geq \eta$, and satisfy **A1** holding with $\delta' = \frac{\delta}{2 N^{1/\log(\frac{1}{1-\eta})}}$ and **A2** of Theorem 4.1.*

The proof of the theorem is given in the Appendix. The constant $k_0$ is specified in Equation 3. Roughly speaking, the theorem implies that under the conditions of the Theorem 4.1 we can robustly recover all clusters of size $O(\log N)$ or larger using only $O(N \log^2 N)$ similarities. Comparing this result to Theorem 3.1, we note three costs associated with being robust to inconsistent similarities: 1) we require $O(N \log^2 N)$ rather than $O(N \log N)$ similarity values; 2) the degree to which the clusters are balanced now plays a role (in the constant $\eta$); 3) we cannot guarantee the recovery of clusters smaller than $O(\log N)$ due to voting.

## 5  Robust Clustering Experiments

To test our robust clustering methodology we focus on experimental results from a balanced binary tree using synthesized similarities and a real-world data set using genetic microarray data ([19]). The synthetic binary tree experiments allows us to observe the characteristics of our algorithm while controlling the amount of inconsistency with respect to the Tight Clustering (TC) condition, while the real world data gives us perspective on problems where the tree structure and TC condition is assumed, but not known.

In order to quantify the performance of the tree reconstruction algorithms, consider the non-unique partial ordering, $\pi : \{1, 2, ..., N\} \to \{1, 2, ..., N\}$, resulting from the ordering of items in the reconstructed tree. For a set of observed similarities, given the original ordering of the items from the true tree structure we would expect to find the largest similarity values clustered around the diagonal of the similarity matrix. Meanwhile, a random ordering of the items would have the large similarity values potentially scattered away from the diagonal. To assess performance of our reconstructed tree structures, we will consider the rate of decay for similarity values off the diagonal of the reordered items, $\widehat{s}_d = \frac{1}{N-d} \sum_{i=1}^{N-d} s_{\pi(i), \pi(i+d)}$. Using $\widehat{s}_d$, we define a distribution over the the average off-diagonal similarity values, and compute the entropy of this distribution as follows:

$$\widehat{E}(\pi) = - \sum_{i=1}^{N-1} \widehat{p}_{\pi_i} \log \widehat{p}_{\pi_i} \qquad (2)$$

Where $\widehat{p}_{\pi_i} = \left( \sum_{d=1}^{N-1} \widehat{s}_d \right)^{-1} \widehat{s}_i$.

This entropy value provides a measure of the quality of a partial ordering induced by the tree reconstruction algorithm. For a balanced binary tree with N=512, we find that for the original ordering, $\widehat{E}(\pi_{original}) = 2.2323$, and for the random ordering, $\widehat{E}(\pi_{random}) = 2.702$. This motivates examining the estimated $\Delta$-entropy of our clustering reconstruction-based orderings as $\widehat{E}_\Delta(\pi) = \widehat{E}(\pi_{random}) - \widehat{E}(\pi)$, where we normalize the reconstructed clustering entropy value with respect to a random permutation of the items. The quality of our clustering methodologies will be examined, where the larger the estimated $\Delta$-entropy, the higher the quality of our estimated clustering.

For the synthetic binary tree experiments, we created a balanced binary tree with 512 items. We generated similarity between each pair of items such that $100 \cdot (1-q)\%$ of the pairwise similarities chosen at random are consistent with the TC condition ($\in \mathbf{S}_C$). The remaining $100 \cdot q\%$ of the pairwise similarities were inconsistent with the TC condition. We examined the performance of both standard bottom-up agglomerative clustering and our Robust Clustering algorithm, `RAcluster`, for pairwise similarities with $q = 0.05, 0.15, 0.25$. The results presented here are averaged over 10 random realization of noisy synthetic data and setting the threshold $\gamma = 0.30$. We used the similarity voting budgets $m = 40$ and $m = 80$, which require 38% and 65% of the complete set of similarities, respectively. Performance gains are shown using our robust clustering approach in Table 2 in terms of both the estimated $\Delta$-entropy and $r_{min}$, the size of the smallest correctly resolved cluster (where all clusters of size $r_{min}$ or larger are reconstructed correctly for the clustering). Comparisons between $\Delta$-entropy and $r_{min}$ show a clear correlation between high $\Delta$-entropy and high clustering reconstruction resolution. While we make no theoretical guarantees for reconstruction when the clusters are smaller than $m$ or when the properties of the clusters $(q, \eta)$ are outside the feasible region of Figure 3, these results show that these clusters can be possibly resolved in practice.

Table 2: Clustering $\Delta$-entropy results for synthetic binary tree with $N = 512$ for Agglomerative Clustering and `RAcluster`.

| $q$ | Agglo. Clustering | | Robust (m=40) | | Robust (m=80) | |
|---|---|---|---|---|---|---|
|  | $\Delta$-Entropy | $r_{min}$ | $\Delta$-Entropy | $r_{min}$ | $\Delta$-Entropy | $r_{min}$ |
| 0.05 | 0.3666 | 460.8 | 1.0178 | 6.8 | 1.0178 | 7.2 |
| 0.15 | 0.0899 | 512 | 1.0161 | 16 | 1.0161 | 15.2 |
| 0.25 | 0.0133 | 512 | 0.9360 | 384 | 1.0119 | 57.6 |

In terms of a real world data set, we test our methodologies against a set of gene microarray data. Our genetic dataset [19] consists of 1,024 yeast genes with 7 expressions each, from which we exhaustively generate the standard Pearson correlation using the expression vectors for every pair of genes. Our robust clustering methodology is performed on the datasets using the threshold $\gamma = 0.30$ and similarity voting budgets $m = 10$ and $m = 80$, requiring at most 18% and 61% of the total similarities (when $N = 512$), respectively. The results in Table 3 (averaged over 10 random permutations of the datasets) show that again our robust clustering methodology outperforms agglomerative clustering in terms of estimated $\Delta$-entropy of the reordered elements, while also not requiring to observe every pairwise similarity value. In Figure 4 we see the reordered similarity matrices given both agglomerative clustering and our robust clustering methodology, `RAcluster`.

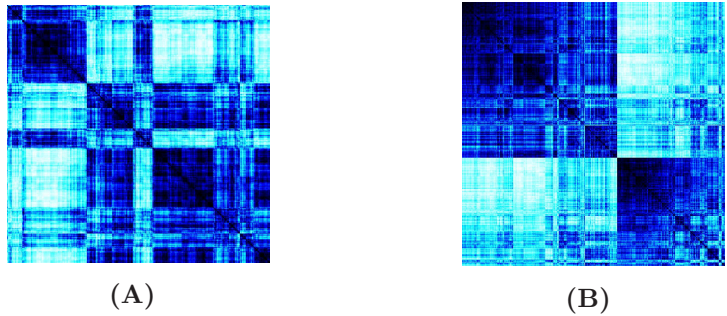

          (A)                                                    (B)

Figure 4: Reordered pairwise similarity matrices, Gene microarray data with $N = 1024$ using (A) - Agglomerative Clustering and (B) - Robust Clustering using $m = 80$ (requiring only 43% of the similarities). An ideal clustering would organize items so that the similarity matrix is dark blue (high similarity) clusters/blocks on the diagonal and light blue (low similarity) values off the diagonal blocks. The robust clustering is clearly closer to this ideal (*i.e.*, B compared to A).

Table 3: $\Delta$-entropy results for real world gene microarray dataset using both Agglomerative Clustering and Robust Clustering algorithms.

| Dataset | Agglo. | Robust (m=10) | Robust (m=80) |
|---|---|---|---|
| Gene (N=512) | 0.1417 | 0.1912 | 0.2035 |
| Gene (N=1024) | 0.0761 | 0.1325 | 0.1703 |

# 6 Conclusions

Despite the wide ranging applications of hierarchical clustering (biology, networking, scientific simulation), relatively little work has been done on examining the number of pairwise similarity values needed to resolve the hierarchical dependencies in the presence of noise. The goal of our work was to use drastically fewer selected pairwise similarity values to reduce the total number of pairwise similarities needed to resolve the hierarchical dependency structure while remaining robust to potential outliers in the data. When there are no outliers, we presented a methodology that requires no more than $3N \log_{\frac{3}{2}} N$ similarity values to recover the clustering hierarchy. We then showed that in the presence of inconsistent similarity values we only require on the order of $O\left(N \log^2 N\right)$ similarities to robustly recover the clustering. These results open hierarchical clustering to a new realm of large-scale problems that were previously impractical to evaluate.

# 7 Appendix

## 7.1 Proof of Theorem 4.1

Since the `outlier` tests can be erroneous, we instead use a two-round voting procedure to correctly determine whether two items $x_i$ and $x_j$ are in the same sub-cluster or not. Please refer to Algorithm 1 for definitions of the relevant quantities. The following lemma establishes that the outlier fraction values $c_{i,k}$ can reveal whether two items $x_i, x_k$ are in the same subcluster or not, provided that the number of voting items $m = |\mathcal{S}_V|$ is large enough and the similarity $s_{i,k}$ is consistent (later we will show that a second round of voting can remove this requirement).

**Lemma 2.** *Consider two items $x_i$ and $x_k$. Under assumptions **A1** and **A2** and assuming $s_{i,k} \in \mathbf{S}_C$, comparing the outlier count values $c_{i,k}$ to a threshold $\gamma$ will correctly indicate whether $x_i, x_k$ are in the same subcluster with probability at least $1 - \frac{\delta_C}{2}$ for*

$$m \geq \frac{\log(4/\delta_C)}{2 \min\left((\gamma - 1 + (1-q)^2)^2, ((1-q)^2\eta - \gamma)^2\right)}.$$

*Proof.* Let $\Omega_{i,k} := \mathbf{1}_{\{s_{i,k} \in \mathbf{S}_C\}}$ be the event that the similarity between items $x_i, x_k$ is in the consistent subset (see Definition 3). Under **A1**, the expected outlier fraction $(c_{i,k})$ conditioned on $x_i, x_k$ and $\Omega_{i,k}$ can be bounded in two cases; when they belong to the same subcluster and when they do not:

$$\mathbb{E}\left[c_{i,k} \mid x_i, x_k \in \mathcal{C}_L \text{ or } x_i, x_k \in \mathcal{C}_R, \Omega_{i,k}\right] \geq (1-q)^2 \eta$$

$$\mathbb{E}\left[c_{i,k} \mid x_i \in \mathcal{C}_R, x_k \in \mathcal{C}_L \text{ or } x_i \in \mathcal{C}_L, x_k \in \mathcal{C}_R, \Omega_{i,k}\right]$$
$$\leq \left(1 - (1-q)^2\right)$$

**A2** stipulates a gap between the two bounds. Hoeffding's Inequality ensures that, with high probability, $c_{i,k}$ will not significantly deviate below/above the lower/upper bound. Thresholding $c_{i,k}$ at a level $\gamma$ between the bounds will probably correctly determine whether $x_i$ and $x_k$ are in the same subcluster or not. More precisely, if

$$m \geq \frac{\log(4/\delta_C)}{2 \min\left((\gamma - 1 + (1-q)^2)^2, ((1-q)^2\eta - \gamma)^2\right)},$$

then with probability at least $(1 - \delta_C/2)$ the threshold test correctly determines if the items are in the same subcluster. □

Next, note that we cannot use the cluster count $c_{i,j}$ directly to decide the placement of $x_i$ since the condition $s_{i,j} \in \mathbf{S}_C$ may not hold. In order to be robust to errors in $s_{i,j}$, we employ a second round of voting based on an independent set of $m$ randomly selected *agreement* items, $\mathcal{S}_A$. The *agreement fraction*, $a_{i,j}$, is the average of the number of times the item $x_i$ agrees with the clustering decision of $x_j$ on $\mathcal{S}_A$.

**Lemma 3.** *Consider the following procedure:*

$$x_i \in \begin{cases} \mathcal{C}_j & : \text{if} \quad a_{i,j} \geq \frac{1}{2} \\ \mathcal{C}_j^c & : \text{if} \quad a_{i,j} < \frac{1}{2} \end{cases}$$

*Under assumptions* **A1** *and* **A2**, *with probability at least* $1 - \frac{\delta_\mathcal{C}}{2}$, *the above procedure based on $m = |\mathcal{S}_A|$ agreement items will correctly determine if the items $x_i, x_j$ are in the same subcluster, provided*

$$m \geq \frac{\log(4/\delta_\mathcal{C})}{2\left((1-\delta_\mathcal{C})(1-q)^2 - \frac{1}{2}\right)^2} \ .$$

*Proof.* Define $\Phi_{i,j}$ as the event that similarities $s_{i,k}, s_{j,k}$ are both consistent (*i.e.*, $s_{i,k}, s_{j,k} \in \mathbf{S}_C$) and thresholding the cluster counts $c_{i,k}, c_{j,k}$ at level $\gamma$ correctly indicates if the underlying items belong to the same subcluster or not. Then using Lemma 2 and the union bound we can bound the probabilities,

$$\begin{aligned} P(\Phi_{i,j}) &\geq (1-\delta_\mathcal{C})(1-q)^2 \\ P(\Phi_{i,j}^C) &\leq 1 - (1-\delta_\mathcal{C})(1-q)^2 \end{aligned}$$

Then the conditional expectations of the agreement counts, $a_{i,j}$, can be bounded as,

$$\begin{aligned} \mathbb{E}\left[a_{i,j} \mid x_i \notin \mathcal{C}_j\right] &\leq P(\Phi_{i,j}^C) \leq 1 - (1-q)^2(1-\delta_C) \\ \mathbb{E}\left[a_{i,j} \mid x_i \in \mathcal{C}_j\right] &\geq P(\Phi_{i,j}) \geq (1-q)^2(1-\delta_C) \end{aligned}$$

Since $q \leq 1 - 1/\sqrt{2(1-\delta')}$ and $\delta_\mathcal{C} = \delta'/n$ (as defined below), there is a gap between these two bounds that includes the value $1/2$. Hoeffding's Inequality ensures that with high probability $a_{i,j}$ will not significantly deviate above/below the upper/lower bound. Thus, thresholding $a_{i,j}$ at $1/2$ will resolve whether the two items $x_i, x_j$ are in the same or different subclusters with probability at least $(1 - \delta_\mathcal{C}/2)$ provided $m \geq \log(4/\delta_\mathcal{C})/\left(2\left((1-\delta_\mathcal{C})(1-q)^2 - \frac{1}{2}\right)^2\right)$. □

By combining Lemmas 2 and 3, we can state the following. The `split` methodology of Algorithm 1 will successfully determine if two items $x_i, x_j$ are in the same subcluster with probability at least $1 - \delta_\mathcal{C}$ under assumption **A1** and **A2**, provided

$$m \geq \max\left(\frac{\log(4/\delta_\mathcal{C})}{2\left((1-\delta_\mathcal{C})(1-q)^2 - \frac{1}{2}\right)^2}, \frac{\log(4/\delta_\mathcal{C})}{2\min\left((\gamma - 1 + (1-q)^2)^2, ((1-q)^2\eta - \gamma)^2\right)}\right)$$

and the cluster under consideration has at least $2m$ items.

In order to successfully determine the subcluster assignments for all $n$ items of the cluster $\mathcal{C}$ with probability at least $1 - \delta'$, requires setting $\delta_C = \frac{\delta'}{n}$ (i.e., taking the union bound over all n items). Thus we have the requirement

$$m \geq c_0\left(\delta', \eta, q, \gamma\right) \log(4n/\delta')$$

where the constant obeys

$$c_0\left(\delta', \eta, q, \gamma\right) \geq \max\left(\frac{1}{2\left(\left(1 - \delta'\right)\left(1 - q\right)^2 - \frac{1}{2}\right)^2}, \frac{1}{2\min\left(\left(\gamma - 1 + (1-q)^2\right)^2, \left((1-q)^2\eta - \gamma\right)^2\right)}\right) \quad (3)$$

Finally, this result and assumptions **A1-A2** imply that the algorithm $\mathtt{split}(\mathcal{C}, m, \gamma)$ correctly determine the two subclusters of $\mathcal{C}$ with probability at least $1 - \delta'$.

### 7.2 Proof of Theorem 4.2

**Lemma 4.** *A binary tree with $N$ leaves and balance factor $\eta_C \geq \eta$ has depth of at most $L \leq \log N / \log(\frac{1}{1-\eta})$.*

*Proof.* Consider a binary tree structure with $N$ leaves (items) with balance factor $\eta \leq 1/2$. After depth of $\ell$, the number of items in the largest cluster are bounded by $(1 - \eta)^\ell N$. If $L$ denotes the maximum depth level, then there can only be 1 item in the largest cluster after depth of $L$, we have $1 \leq (1 - \eta)^L N$. □

The entire hierarchical clustering can be resolved if all the clusters are resolved correctly. With a maximum depth of $L$, the total number of clusters $M$ in the hierarchy is bounded by $\sum_{\ell=0}^{L} 2^\ell \leq 2^{(L+1)} \leq 2N^{1/\log(\frac{1}{1-\eta})}$, using the result of Lemma 4. Therefore, the probability that some cluster in the hierarchy is not resolved $\leq M\delta' \leq 2N^{1/\log(\frac{1}{1-\eta})}\delta'$ (where $\mathtt{split}$ succeeds with probability $> 1 - \delta'$). Therefore, for all clusters (which satisfy the conditions **A1** and **A2** of Theorem 4.1 and have size $> 2m$) can be resolved with probability $1 - \delta$ (by setting $\delta' = \frac{\delta}{2N^{1/\log(\frac{1}{1-\eta})}}$), from the proof of Theorem 4.1, we require,

$$m \geq \max\left(\frac{\log\left(\frac{8}{\delta}N^{1+(1/\log(\frac{1}{1-\eta}))}\right)}{2\left(\left(1 - \frac{\delta}{2N^{1+(1/\log(\frac{1}{1-\eta}))}}\right)^2 (1-q)^2 - \frac{1}{2}\right)^2}, \frac{\log\left(\frac{8}{\delta}N^{1+(1/\log\frac{1}{1-\eta})}\right)}{2\min\left(\left(\gamma - 1 + (1-q)^2\right)^2, \left((1-q)^2\eta - \gamma\right)^2\right)}\right)$$

which holds if $m = k_0(\delta, \eta, q, \gamma) \log\left(\frac{8}{\delta} N\right)$ where,

$$k_0(\delta, \eta, q, \gamma) \geq c_0(\delta, \eta, q, \gamma) / \left(1 + \left(1/\log\left(\frac{1}{1-\eta}\right)\right)\right) \quad (4)$$

Given this choice of $m$, we find that the `RAcluster` methodology in Algorithm 3 for a set of $N$ items will resolve all clusters that satisfy **A1** and **A2** of Theorem 4.1 and have size $> 2m$, with probability at least $1 - \delta$.

Furthermore, the algorithm only requires $O\left(N \log^2 N\right)$ total pairwise similarities. By running the `RAcluster` methodology, using the result from Lemma 4, each item will have the `split` methodology performed at most $\log N / \log(\frac{1}{1-\eta})$ times (*i.e.,* once for each depth level of the hierarchy). If $m = k_0(\delta, \eta, q, \gamma) \log\left(\frac{8}{\delta} N\right)$ for `RAcluster`, each call to `split` will require only $3k_0(\delta, \eta, q, \gamma) \log\left(\frac{8}{\delta} N\right)$ pairwise similarities per item. Given $N$ total items, we find that the `RAcluster` methodology requires at most $3k_0(\delta, \eta, q, \gamma) N \frac{\log N}{\log(\frac{1}{1-\eta})} \log\left(\frac{8}{\delta} N\right)$ pairwise similarities.